\documentclass[pdflatex,sn-mathphys-num]{sn-jnl}

\usepackage{graphicx}%
\usepackage{multirow}%
\usepackage{amsmath,amssymb,amsfonts}%
\usepackage{amsthm}%
\usepackage{mathrsfs}%
\usepackage[title]{appendix}%
\usepackage{xcolor}%
\usepackage{textcomp}%
\usepackage{manyfoot}%
\usepackage{booktabs}%
\usepackage{algorithm}%
\usepackage{algorithmicx}%
\usepackage{algpseudocode}%
\usepackage{listings}%

\theoremstyle{thmstyleone}%
\theoremstyle{thmstyletwo}%

\theoremstyle{thmstylethree}%
\newtheorem{definition}{Definition}%

\begin{document}

\title[Article Title]{Optimal convex approximation of quantum channels based on $\alpha$-affinity}


\author[1]{\fnm{Liqiang} \sur{Zhang}}\email{zhangliqiang@sxnu.edu.cn}

\author[1]{\fnm{Chengling} \sur{Fu}}\email{2746836190@qq.com}

\author[1]{\fnm{Liuyong} \sur{Cheng}}\email{lycheng@sxnu.edu.cn}

\author*[2]{\fnm{Guohui} \sur{Yang}}\email{yangguohui\_1981\_1981@126.com}

\author*[1]{\fnm{Changshui} \sur{Yu}}\email{ycs@dlut.edu.cn}

\affil*[1]{\orgdiv{School of Physics and Electronic Engineering}, \orgname{Shanxi Normal University}, \orgaddress{\city{Taiyuan}, \postcode{030031}, \country{China}}}
\affil[2]{\orgdiv{School of Physics}, \orgname{Dalian University of Technology}, \orgaddress{\city{Dalian}, \postcode{116024}, \country{China}}}


\abstract{Determining the minimal distance between a target channel and a convex hull of predefined set of implementable channels is a fundamental problem in quantum resource theory, and provides key guidance for experimental implementations. In this work, we develop a unified analytical framework for optimal convex approximation of quantum channels based on the quantum $\alpha$-affinity measure. We construct a channel distance metric induced by the $\alpha$-affinity and the Choi–Jamiolkowski isomorphism, which satisfies the required properties of a well-defined channel distance. Subsequently, we present an optimization framework for the convex approximation of quantum channels, and derive analytical solutions for the optimal convex approximation of single-qubit unitary channels over both the SU(2)-covariant and Pauli channel families, obtaining closed-form expressions for the optimal parameters and the minimal approximation distance. This framework is further applied to the amplitude-damping channel, yielding the explicit form of its optimal approximation and the associated minimal $\alpha$-affinity distance. In contrast to conventional approaches based on the diamond norm, our framework provides a systematic and analytically tractable approach to quantum channel approximation under realistic constraints.}




\maketitle

\section{Introduction}\label{sec1}

Quantum resource theory (QRT)~\cite{R1,R2,R3,R4} has emerged as one of the central paradigms in quantum information science, aiming to characterize and quantify the nonclassical properties of quantum systems~\cite{R5}. Within this framework, various quantum features, including entanglement~\cite{R6,R7,R8,R9,R10,R11,R12,R13,R14,R15,R16,R17,R18,R19}, coherence~\cite{R20,R21,R22,R23,R24,R25,R26,R27,R28,R29,R30,R31}, asymmetry~\cite{R32,R33,R34}, and quantum correlations~\cite{R35,R36,R37,R38} etc., can be systematically analyzed through the notions of free states, free operations, and resource measures. A fundamental problem in quantum resource measure theories is to determine how accurately a target quantum state can be approximated by a convex combination of free states. Convex approximation methods have therefore been widely investigated in the contexts of entanglement, quantum coherence, and other resource theories, both from a foundational perspective~\cite{R39} and within the framework of operational resource theories. Representative examples include the best separable approximation of entangled states~\cite{R41} and the optimal approximation of quantum states by incoherent or classical states, which have been studied using convex optimization techniques and R\'enyi-type divergences~\cite{R42,R43}. Therefore, optimally approximating a target quantum state via convex combinations of a set of accessible quantum states not only helps refine quantum resource theories, but also carries practical guiding significance for experiments ~\cite{R44,R45,R46,R47,R48,R49,R50,R51}.

Compared with the well-developed theory for quantum states, the convex approximation problem for quantum channels remains much less explored. Quantum channels~\cite{R52,R53} describe the most general physical evolution of open quantum systems and play a central role in quantum communication, quantum computation, and quantum error correction. For a given inaccessible quantum channel $\Phi$, the channel convex approximation problem is to find the most indistinguishable channel from the accessible channels $\{\Phi_i\}$, such that the distance between $\Phi$ and convex set $\sum_i p_i \Psi_i$ is minimized. Existing studies on channel convex approximation mainly rely on operational distance measures such as the diamond norm~\cite{R54,R55}. Although such distances possess clear operational interpretations in channel discrimination~\cite{R56,R57,R58,R59,R60,R61,R62,R63,R64} tasks, their optimization is often analytically difficult and usually requires extensive numerical computations. Consequently, obtaining explicit analytical solutions for optimal channel approximations remains highly challenging~\cite{R55,R65}. This motivates the search for alternative channel distance measures that simultaneously possess good mathematical properties and analytical tractability~\cite{R66,R67}.

A typical derived problem lies in the selection of quantum channel distance measures. In this work we establish a unified framework for the optimal convex approximation of quantum channels based on the $\alpha$-affinity measure~\cite{R68,R69}. The $\alpha$-affinity has attracted increasing attention in quantum information theory due to its close relation to Rényi-type quantities and its favorable mathematical properties, including joint concavity and monotonicity under completely positive trace-preserving maps~\cite{R70,R71}. The $\alpha$-affinity has been successfully applied to the quantification of quantum coherence~\cite{R72}. We then analytically study the optimal convex approximation of several representative single-qubit quantum channels. For each case, we derive explicit closed-form expressions for the optimal convex weights and minimal approximation distances.\par
In this work, we focus on the optimal convex approximation of quantum channels based on the $\alpha$-affinity metric, and systematically investigate the induced channel distance for several typical single-qubit channels. The organization of the paper is as follows. In Sec.~2, we derive the distance measure between channels induced by the $\alpha$-affinity and prove its fundamental mathematical properties, such as contractivity under superchannels, and then establish an optimization framework for the optimal convex approximation of quantum channels using the $\alpha$-affinity metric. In Sec.~3, we investigate the optimally covariant convex approximation of unitary channels. In Sec.~4 and 5, we study the optimal Pauli convex approximation of unitary channels separately. Finally, we present our conclusions In Sec.~6.
\section{Optimization Framework Based on $\alpha$-Affinity}\label{sec2}
\subsection{Quantum Channel Distance Measure Based on $\alpha$-Affinity}\label{subsec2-1}
In quantum information theory, quantum channel is a completely positive and trace-preserving linear map that describes the dynamical evolution of quantum states. Let $\mathcal{H}$ be a finite-dimensional Hilbert space, and let $B(\mathcal{H})$ denote the set of density operators on $\mathcal{H}$. A quantum channel $\Phi: B(\mathcal{H}_A) \to B(\mathcal{H}_B)$ is defined as a completely positive and trace-preserving (CPTP) linear map between operator spaces.

In practice, it is often impossible to exactly implement a target channel $\Phi$; instead, one may only have access to a set of implementable channels $\{\Psi_i\}$. A natural problem therefore arises: how to approximate the target channel using convex combinations of these available channels, and how to quantify the approximation error. To address this, we introduce a suitable state distance measure based on the $\alpha$-affinity ($0 < \alpha < 1$) as follows \cite{R73,R74}:
\begin{equation}
M_\alpha(\rho,\sigma) = 1 - A_\alpha(\rho,\sigma).
\end{equation}
where $\mathcal{A}_\alpha(\rho, \sigma) = \operatorname{Tr}\left( \rho^\alpha \sigma^{1-\alpha} \right)$. This definition quantifies the dissimilarity between quantum states: when two states are close, their $\alpha$-affinity approaches one, and the corresponding distance tends to zero; conversely, when the states are more distinguishable, the $\alpha$-affinity decreases and the distance increases.\par
To extend the $\alpha$-affinity from quantum states to quantum channels, we employ the Choi--Jamiolkowski isomorphism. For a quantum channel $\Phi(\cdot)$, its normalized Choi operator is defined as~\cite{R72,R74}
\begin{equation}
R_\Phi = \frac{J_\Phi}{d} =  \big( \Phi \otimes \mathbb{I} \big) \vert \eta \rangle \langle \eta \vert,
\end{equation}
where $|\eta\rangle = \frac{1}{\sqrt{d}} \sum_{i=0}^{d-1}|i\rangle \otimes |i\rangle$ is a maximally entangled state in $\mathcal{H}_A \otimes \mathcal{H}_B$, and $\mathbb{I}$ is the identity operator. We consider $ R_\Phi$ is a valid quantum state. In the following, all channel distances are defined in terms of the normalized Choi states.\par
Next, we define a distance metric for quantum channels via the Choi–Jamiolkowski isomorphism and $\alpha$-affinity.
\begin{definition}[Definition of $\alpha$-channel distance]
Given two quantum channels $\Phi$ and $\Psi$ with the corresponding Choi states $R_\Phi$ and $R_\Psi$, the $\alpha$-distance between these two channels is defined as
\begin{equation}
\begin{aligned}
D_\alpha(\Phi,\Psi)= M_\alpha\left( \frac{J_\Phi}{d},\ \frac{J_\Psi}{d} \right) = 1 - \mathcal{A}_\alpha\left(\frac{J_\Phi}{d},\ \frac{J_\Psi}{d}\right) = 1 - \mathcal{A}_\alpha(R_\Phi, R_\Psi).
\end{aligned}
\label{eq:alpha_distance}
\end{equation}
where $R_\Phi = \dfrac{J_\Phi}{d}$ and $R_\Psi = \dfrac{J_\Psi}{d}$.
\end{definition}
The channel $\alpha$-distance $D_\alpha(\Phi,\Psi)$ is a well-defined distance measure for quantum channels, which satisfies the following properties:
\begin{enumerate}
    \item[($P1$)] Positivity: $D_\alpha(\Phi,\Psi) \ge 0$, and $D_\alpha(\Phi,\Psi) = 0$ if and only if $\Phi = \Psi$.
    \item[($P2$)] Contractivity under superchannels: 
For any superchannel $\mathcal{E}$,
   \begin{equation}
    D_\alpha\big(\mathcal{E}(\Phi), \mathcal{E}(\Psi)\big) \le D_\alpha(\Phi,\Psi).
    \end{equation}
    \item[($P3$)] Joint convexity: For any probability distribution $\{p_i\}$ satisfying $p_i \ge 0$ and $\sum_i p_i = 1$,
    \begin{equation}
    D_\alpha\left(\sum_i p_i \Phi_i, \sum_i p_i \Psi_i\right)
    \le \sum_i p_i D_\alpha(\Phi_i, \Psi_i).
    \end{equation}
\end{enumerate}

\begin{proof}
($P1$) Positivity. 
First, recall that the $\alpha$-affinity $\mathcal{A}_\alpha(\rho,\sigma)$ for any two quantum states $\rho$ and $\sigma$ satisfies the inequality
\begin{equation}
0 \le \mathcal{A}_\alpha(\rho,\sigma) \le 1.
\end{equation}
Since $R_\Phi$ and $R_\Psi$ are normalized Choi states, substituting them into the definition of the $\alpha$-affinity, we obtain
\begin{equation}
0 \le \mathcal{A}_\alpha(R_\Phi, R_\Psi) \le 1.
\end{equation}
According to the definition of the channel $\alpha$-distance
\begin{equation}
\begin{aligned}
D_\alpha(\Phi,\Psi)= M_\alpha\left( \frac{J_\Phi}{d},\ \frac{J_\Psi}{d} \right) = 1 - \mathcal{A}_\alpha\left( \frac{J_\Phi}{d},\ \frac{J_\Psi}{d} \right) = 1 - \mathcal{A}_\alpha(R_\Phi, R_\Psi) \ge 0,
\end{aligned}
\end{equation}
which proves the non-negativity part.

Next, we prove the equality condition (identity of indiscernibles):
\begin{itemize}
    \item If $\Phi = \Psi$, then their Choi operators satisfy $J_\Phi = J_\Psi$, so the normalized Choi states are equal: $R_\Phi = R_\Psi$. For any quantum state $\rho$, $\mathcal{A}_\alpha(\rho,\rho) = 1$, so $D_\alpha(\Phi,\Psi) = 0$.
    \item Conversely, if $D_\alpha(\Phi,\Psi) = 0$, then $1 - \mathcal{A}_\alpha(R_\Phi, R_\Psi) = 0$, i.e., $\mathcal{A}_\alpha(R_\Phi, R_\Psi) = 1$. By the property of the $\alpha$-affinity: $\mathcal{A}_\alpha(\rho,\sigma) = 1$ if and only if $\rho = \sigma$, thus $R_\Phi = R_\Psi$. Since the Choi–Jamiolkowski isomorphism is bijective, $R_\Phi = R_\Psi$ implies $J_\Phi = J_\Psi$, and further $\Phi = \Psi$.
\end{itemize}
Combining both parts, $D_\alpha(\Phi,\Psi) \ge 0$ and $D_\alpha(\Phi,\Psi) = 0 \iff \Phi = \Psi$, the positivity property holds.

($P2$) Contractivity under superchannels.
Let $\mathcal{E}$ be an arbitrary superchannel acting on quantum channels. By the definition of superchannels, they correspond to completely positive trace-preserving (CPTP) maps acting on the normalized Choi states of the channels. That is, the normalized Choi state of $\mathcal{E}(\Phi)$ is $\mathcal{E}(R_\Phi)$, and the normalized Choi state of $\mathcal{E}(\Psi)$ is $\mathcal{E}(R_\Psi)$, where $\mathcal{E}$ is a CPTP map.

Recall that the $\alpha$-affinity is monotonic under CPTP maps: for any CPTP map $\mathcal{E}$ and quantum states $\rho, \sigma$,
\begin{equation}
\mathcal{A}_\alpha(\mathcal{E}(\rho), \mathcal{E}(\sigma)) \ge \mathcal{A}_\alpha(\rho, \sigma).
\end{equation}
Substituting $R_\Phi$ and $R_\Psi$ into this inequality, we get
\begin{equation}
\mathcal{A}_\alpha(\mathcal{E}(R_\Phi), \mathcal{E}(R_\Psi)) \ge \mathcal{A}_\alpha(R_\Phi, R_\Psi).
\end{equation}
After a simple mathematical transformation, we obtain
\begin{equation}
1 - \mathcal{A}_\alpha(\mathcal{E}(R_\Phi), \mathcal{E}(R_\Psi)) \le 1 - \mathcal{A}_\alpha(R_\Phi, R_\Psi).
\end{equation}
By the definition of $D_\alpha$, the above formula can be further rewritten as
\begin{equation}
D_\alpha(\mathcal{E}(\Phi), \mathcal{E}(\Psi)) \le D_\alpha(\Phi,\Psi),
\end{equation}
proving the contractivity under superchannels.

($P3$) Joint convexity.
 For any probability quantum states $\{\rho_i\}, \{\sigma_i\} and $weights $\{p_i\}$, the joint convexity of $\alpha$-affinity can be expressed as
\begin{equation}
\mathcal{A}_\alpha\left( \sum_{i} p_i \rho_i, \sum_{i} p_i \sigma_i \right) \ge \sum_{i} p_i \mathcal{A}_\alpha(\rho_i, \sigma_i).
\end{equation}
Substituting $\rho_i = R_{\Phi_i}$ and $\sigma_i = R_{\Psi_i}$, we get
\begin{equation}
\mathcal{A}_\alpha\left( \sum_{i} p_i R_{\Phi_i}, \sum_{i} p_i R_{\Psi_i} \right) \ge \sum_{i} p_i \mathcal{A}_\alpha(R_{\Phi_i}, R_{\Psi_i}).
\end{equation}
It can be further rewritten as
\begin{equation}
1 - \mathcal{A}_\alpha\left( \sum_{i} p_i R_{\Phi_i}, \sum_{i} p_i R_{\Psi_i} \right) \le 1 - \sum_{i} p_i \mathcal{A}_\alpha(R_{\Phi_i}, R_{\Psi_i}).
\end{equation}
Combining with the definition of $D_\alpha$, we obtain
\begin{equation}
D_\alpha\left( \sum_{i} p_i \Phi_i, \sum_{i} p_i \Psi_i \right) \le \sum_{i} p_i D_\alpha(\Phi_i, \Psi_i),
\end{equation}
which proves the joint convexity.
$D_\alpha(\Phi,\Psi)$ satisfies all the properties and is a well-defined quantum channel distance measure.
\end{proof}
\subsection{Convex Approximation of quantum channel}\label{subsec2-2}
Next, we present the definition of optimal convex approximation for quantum channels based on $\alpha$-affinity as follows.
\begin{definition}[Optimal Convex Approximation of Quantum Channels via $\alpha$-affinity]
Given a target channel $\Phi$ and a set of available channels $\{\Psi_i\}$, the convex approximation of the quantum channel $\Phi$ via the $\alpha$-affinity is defined as follows:
\begin{equation}
\widetilde{D}(\Phi,\Psi) = \min_{\{p_i\}} D_\alpha\left(\Phi,\sum_i p_i \Psi_i\right),
\end{equation}
where $\{p_i\}$ is a probability distribution satisfying $p_i \ge 0$ and $\sum_i p_i = 1$.
The optimal convex approximation of a quantum channel $\Phi$ with respect to a given set of quantum channels $\{\Psi_i\}$ is given by $\sum_i p_i^{\text{opt}} \Psi_i$, where $\{p_i^{\text{opt}}\}$ denotes the vector of optimal probabilities, defined as
\begin{equation}
\{p_i^{\text{opt}}\} = \arg\min_{\{p_i\}} \left[1 - \mathcal{A}_\alpha\left(\Phi, \sum_i p_i \Psi_i\right)\right],
\end{equation}
The effectiveness of the optimal convex approximation is quantified by the $\{\Psi_i\}$-distance, which is given by
\begin{equation}
\widetilde{D}_{\{\Psi_i\}}(\Phi) = \min_{\{p_i\}} \left[1 - \mathcal{A}_\alpha\left(\Phi, \sum_i p_i \Psi_i\right)\right].
\end{equation}
\end{definition}
\section{Distance of a Unitary Map and Covariant Channels}\label{sec3}
Let us now study the case of qubits, where the target map $\Phi$ is a unitary transformation. Up to a global phase, it can be parameterized as
\begin{equation}
U(\gamma,\beta,\delta)=
\begin{pmatrix}
\cos\gamma \,e^{i\beta} & \sin\gamma \,e^{i\delta} \\
-\sin\gamma \,e^{-i\delta} & \cos\gamma \,e^{-i\beta}
\end{pmatrix},
\end{equation}
where the parameter space is given by
$\gamma\in\left[0,\frac{\pi}{2}\right],\beta\in[0,2\pi]$, and $\delta\in[0,2\pi].$ The corresponding unitary channel reads
\begin{equation}
\mathcal{U}(\rho)=U\rho U^\dagger,
\end{equation}
which represents an ideal noiseless evolution and serves as the target channel in quantum computation tasks. The Choi matrix of a unitary channel is a pure state given by
\begin{equation}
R_{\mathcal{U}}=(\mathcal{U}\otimes I)|\eta\rangle\langle\eta|=|\psi_U\rangle\langle\psi_U|,
\end{equation}
Expanding $|\psi_U\rangle$ in the Bell basis $\{|B_0\rangle,|B_1\rangle,|B_2\rangle,|B_3\rangle\}$ yields
\begin{equation}
|\psi_U\rangle=a_0|B_0\rangle+a_1|B_1\rangle+a_2|B_2\rangle+a_3|B_3\rangle,
\end{equation}
where the Bell basis reads
\begin{equation}
\begin{aligned}
|B_0\rangle &= \tfrac{1}{\sqrt{2}}(|00\rangle+|11\rangle), |B_1\rangle = \tfrac{1}{\sqrt{2}}(|00\rangle-|11\rangle),\\
|B_2\rangle &= \tfrac{1}{\sqrt{2}}(|01\rangle+|10\rangle), |B_3\rangle = \tfrac{1}{\sqrt{2}}(|01\rangle-|10\rangle).
\end{aligned}
\end{equation}
And with expansion coefficients
\begin{equation}
\begin{aligned}
a_0=\cos\gamma\cos\beta,
a_1=i\sin\gamma\sin\delta,\\
a_2=\sin\gamma\cos\delta,
a_3=i\cos\gamma\sin\beta.
\end{aligned}
\end{equation}
Since $R_{\mathcal{U}}$ is a pure state, it satisfies the idempotent property for fractional powers:
\begin{equation}
R_{\mathcal{U}}^\alpha=R_{\mathcal{U}}.
\end{equation}
\par
Next, we regard covariant channels as accessible channels:
\begin{equation}
\mathcal{C}_p(\rho)=(1-p)\rho+\frac{p}{3}\sum_{i=1}^3 \sigma_i \rho \sigma_i,\quad p\in[0,1],
\end{equation}
it is the convex hull spanned by the identity channel and the equally weighted Pauli rotation channels, satisfying the SU(2)-covariance condition~\cite{R55}:
\begin{equation}
U_g^\dagger \mathcal{C}_p\left(U_g \rho U_g^\dagger\right)U_g = \mathcal{C}_p(\rho),\quad \forall U_g\in\mathrm{SU}(2).
\end{equation}
The Choi matrix of the covariant channel $\mathcal{C}_\mathcal{P}$ is diagonal in the Bell basis:
\begin{equation}
R_{\mathcal{C}_p}=(1-p)|\eta\rangle\langle\eta|+\frac{p}{3}\sum_{i=1}^3 |B_i\rangle\langle B_i|,
\end{equation}
with eigenvalues
$\lambda_0=1-p$ and $\lambda_1=\lambda_2=\lambda_3=\frac{p}{3}$. Accordingly, the $(1-\alpha)$-th power of $R_{\mathcal{C}_p}$ is the diagonal matrix
\begin{equation}
R_{\mathcal{C}_p}^{1-\alpha} =
\begin{pmatrix}
(1-p)^{1-\alpha} & 0 & 0 & 0 \\
0 & \left(\dfrac{p}{3}\right)^{1-\alpha} & 0 & 0 \\
0 & 0 & \left(\dfrac{p}{3}\right)^{1-\alpha} & 0 \\
0 & 0 & 0 & \left(\dfrac{p}{3}\right)^{1-\alpha}
\end{pmatrix} .
\end{equation}
Substituting the Choi matrices into the definition of $\alpha$-affinity, we obtain
\begin{equation}
\begin{aligned}
\mathcal{A}_\alpha(R_{\mathcal{U}},R_{\mathcal{C}_p})=\mathrm{Tr}\!\left(R_{\mathcal{U}}^\alpha R_{\mathcal{C}_p}^{1-\alpha}\right)=\langle\psi_U|R_{\mathcal{C}_p}^{1-\alpha}|\psi_U\rangle=|a_0|^2 (1-p)^{1-\alpha}+\left(\sum_{i=1}^3 |a_i|^2\right)\left(\frac{p}{3}\right)^{1-\alpha}.
\end{aligned}
\end{equation}
The $\alpha$-affinity is then simplified to a univariate function of $p$ :
\begin{equation}
\mathcal{A}_\alpha(p)=|a_0|^2(1-p)^{1-\alpha}+(1-|a_0|^2)\left(\frac{p}{3}\right)^{1-\alpha}.
\end{equation}
\par
\begin{figure*}[tb]
\centering
\begin{minipage}[t]{0.48\textwidth}
\centering
\includegraphics[width=\textwidth]{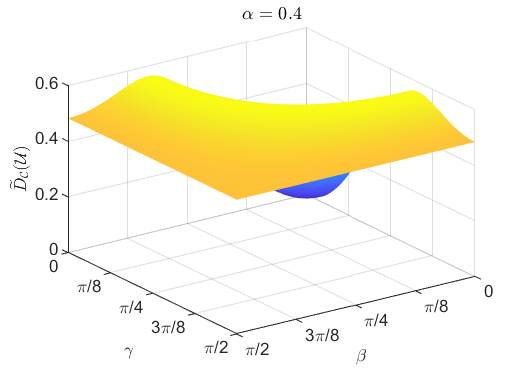}
\vspace{-0.5em}
(a)
\end{minipage}
\hfill
\begin{minipage}[t]{0.48\textwidth}
\centering
\includegraphics[width=\textwidth]{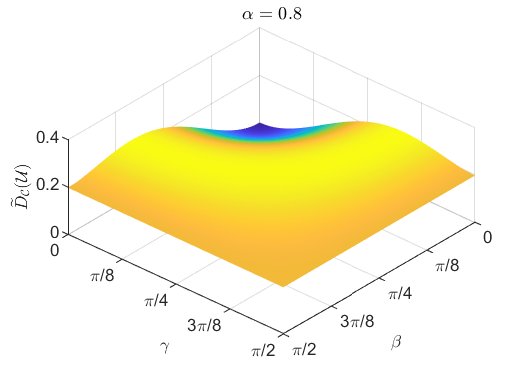}
\vspace{-0.5em}
(b)
\end{minipage}

\vspace{1em} 

\begin{minipage}[t]{0.48\textwidth}
\centering
\includegraphics[width=\textwidth]{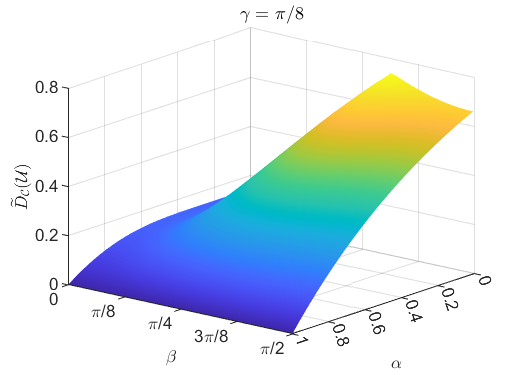}
\vspace{-0.5em}
(c)
\end{minipage}
\hfill
\begin{minipage}[t]{0.48\textwidth}
\centering
\includegraphics[width=\textwidth]{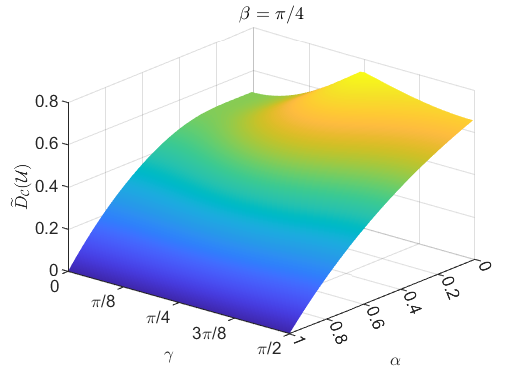}
\vspace{-0.5em}
(d)
\end{minipage}

\caption{Optimal convex approximation of a single-qubit unitary channel $\mathcal{U}(\gamma,\beta,\delta)$ w.r.t. the covariant channel set under the $\alpha$-affinity metric. The problem is solved by finding the closest covariant channel to the target unitary channel, yielding the optimal covariant distance $\widetilde{D}_{\mathcal{C}}[\mathcal{U}(\gamma,\beta,\delta)]$. The solution provides the optimal distance, here plotted versus different pairs of parameters under various fixed conditions: (a) varying $\beta$ and $\gamma$ for a fixed value of $\alpha$, namely, $\alpha=0.4$; (b) varying $\beta$ and $\gamma$ for a fixed value of $\alpha$, namely, $\alpha=0.8$; (c) varying $\alpha$ and $\beta$ for a fixed value of $\gamma$, namely, $\gamma\equiv\pi/8$; (d) varying $\alpha$ and $\gamma$ for a fixed value of $\beta$, namely, $\beta\equiv\pi/4$.}
\label{fig:optimal_distance}
\end{figure*}
As shown earlier, the optimal convex approximation problem is equivalent to maximizing the $\alpha$-affinity between the target unitary channel and the covariant channel family. The problem thus reduces to maximizing $\mathcal{A}_\alpha(p)$ over $p \in [0,1]$. Next, we take the derivative of $\mathcal{A}_\alpha(p)$ with respect to $p$ and obtain:
\begin{equation}
\frac{d\mathcal{A}_\alpha}{dp} = -(1-\alpha)|a_0|^2(1-p)^{-\alpha} + (1-\alpha)\left(1-|a_0|^2\right)\frac{1}{3^{1-\alpha}}p^{-\alpha}.
\end{equation}
Setting this derivative equal to zero yields the algebraic equation for $p$, whose solution is the optimal parameter $p^{\text{opt}}$.\par
When $0 < |a_0|^2 < 1$, the optimal convex weight is solved as
\begin{equation}
p^{\text{opt}} = \frac{(1-|a_0|^2)^{\frac{1}{\alpha}}}{|a_0|^{\frac{2}{\alpha}} \cdot 3^{\frac{1-\alpha}{\alpha}} + (1-|a_0|^2)^{\frac{1}{\alpha}}}.
\end{equation}
Substituting $p^{\text{opt}}$ into the distance formula yields the minimal $\alpha$-distance from the unitary channel to the covariant channel set:
\begin{equation} \label{EEE}
\widetilde{D}_{\mathcal{C}}(\mathcal{U}) = 1 - \left[ (1 - |a_0|^2)^{1/\alpha} + |a_0|^{2/\alpha} \, 3^{\frac{1-\alpha}{\alpha}} \right]^\alpha.
\end{equation}
Equation (\ref{EEE}) gives the closed-form expression for the minimal 
$\alpha$-affinity covariant distance between the target unitary 
channel and the covariant channel set. The result explicitly 
depends on the dominant Bell-basis coefficient $|a_0|^2$ and the 
optimal probability parameter $p^{\mathrm{opt}}$, showing how the 
structure of the target unitary channel determines the achievable 
approximation accuracy. \par
Next, we discuss the boundary and special cases.\par
(1) When $|a_0|^2 = 1$, $\widetilde{D}_{\mathcal{C}}(\mathcal{U}) = 1 - (1-p)^{1-\alpha}.$  Since $(1-p)^{1-\alpha}$ is strictly increasing as $p$ decreases, $\widetilde{D}_{\mathcal{C}}(\mathcal{U})$ is strictly increasing in $p$. Thus, the minimum is attained at $p^{\mathrm{opt}}=0$, yielding $\widetilde{D}_{\mathcal{C}}(\mathcal{U}) = 0.$\par
(2) When $|a_0|^2 = 0$, $\widetilde{D}_{\mathcal{C}}(\mathcal{U}) = 1 - \left( \frac{p}{3} \right)^{1-\alpha}.$ Since $\left( \frac{p}{3} \right)^{1-\alpha}$ is strictly increasing in $p$, $\widetilde{D}_{\mathcal{C}}$ is strictly decreasing in $p$. Thus, the minimum is attained at $p^{\mathrm{opt}}=1$, yielding $\widetilde{D}_{\mathcal{C}}(\mathcal{U}) = 1 - 3^{-(1-\alpha)}.$

Figure 1 presents the comprehensive results of the optimal covariant approximation distance $\widetilde{D}_{\mathcal{C}}(\mathcal{U})$ for single-qubit unitary channels under the $\alpha$-affinity metric.
Figure 1(a) plots the distance landscape with respect to unitary parameters $\beta$ and $\gamma$ at a smaller affinity parameter $\alpha=0.4$, where the distance distribution exhibits higher sensitivity to parameter variations due to the stronger weighting effect induced by small $\alpha$ in the fractional-power structure of the $\alpha$-affinity.
Figure 1(b) shows the distance behavior at $\alpha=0.8$, and the approximation distance strongly depends on the unitary parameters $\gamma$ and $\beta$, whose nontrivial geometric structure reflects the competition among different Bell-basis components in the optimal convex approximation, with larger deviations from the dominant Bell component generally leading to larger approximation distances.
Figure 1(c) depicts the optimal distance with fixed $\gamma=\pi/8$, where the distance mainly depends on the phase parameter $\beta$ that determines the relative contribution of different Bell components in the unitary channel. The periodic variation reflects the intrinsic symmetry of the SU(2) parameterization, and the approximation error remains relatively smooth over the entire parameter region, demonstrating the stability of the $\alpha$-affinity based optimization framework under phase variations.
Figure 1(d) illustrates the distance as functions of the unitary parameter $\gamma$ and the affinity parameter $\alpha$. The approximation error varies smoothly with both parameters, indicating that the $\alpha$-affinity distance provides a stable characterization of the distinguishability between the target unitary channel and the covariant channel family, and the distance increases as the unitary channel deviates further from the identity component in the Bell basis decomposition.
Overall, the results confirm that the affinity parameter $\alpha$ plays a crucial role in controlling and characterizing the distinguishability between the target unitary channel and the covariant channel family.
\section{Pauli Distance of Unitary Channel}\label{sec4}
In this section, we continue to investigate the optimal convex approximation of single-qubit unitary channels under the $\alpha$-affinity metric. Different from the SU(2)-covariant channel family employed previously, we here utilize the full Pauli channel set as the approximation set. Since the parameterization of unitary channels, the Bell-state expansion, and the fundamental definitions of $\alpha$-affinity have been presented earlier, we only focus on the construction of Pauli channels, the optimization of convex weights, and the derivation of the minimal distance.
\par
The approximation set is given by the convex hull of four Pauli channels, which forms a widely used model for noisy quantum operations. A general Pauli channel can be written as
\begin{equation}
\mathcal{P}(\rho) = \sum_{i=0}^{3} p_i \sigma_i \rho \sigma_i,
\end{equation}
where $p_i \ge 0$ and $\sum_{i=0}^{3} p_i = 1$ are convex combination probabilities, and $\{\sigma_0=I, \sigma_1=\sigma_x, \sigma_2=\sigma_y, \sigma_3=\sigma_z\}$ are the single-qubit Pauli matrices.
The Choi matrix of the Pauli channel is diagonal in the Bell basis $\{|B_0\rangle,|B_1\rangle,|B_2\rangle,|B_3\rangle\}$, and can be expressed as
\begin{equation}
R_{\mathcal{P}} = \sum_{i=0}^{3} p_i |B_i\rangle\langle B_i|,
\end{equation}
For any $0<\alpha<1$, its $(1-\alpha)$-th power is
\begin{equation}
R_{\mathcal{P}}^{1-\alpha} = \sum_{i=0}^{3} p_i^{1-\alpha} |B_i\rangle\langle B_i|,
\end{equation}
Substituting the Choi matrices into the definition of $\alpha$-affinity, we obtain
\begin{equation}
\begin{aligned}
\mathcal{A}_\alpha(\mathcal{U},\mathcal{P})= \mathrm{Tr}\!\left( R_{\mathcal{U}}^\alpha R_{\mathcal{P}}^{1-\alpha} \right) = \langle \psi_U | R_{\mathcal{P}}^{1-\alpha} | \psi_U \rangle = \sum_{i=0}^{3} |a_i|^2 \, p_i^{1-\alpha},
\end{aligned}
\end{equation}
The optimal approximation is equivalent to maximizing the $\alpha$-affinity under the normalization constraint
\begin{equation}
\max_{p_i} \sum_{i=0}^{3} |a_i|^2 p_i^{1-\alpha}
\quad \text{s.t.} \quad
\sum_{i=0}^{3} p_i = 1, p_i \geq 0.
\end{equation}
We construct the Lagrangian function~\cite{R49,R75}
\begin{equation}
\mathcal{L} = \sum_{i=0}^{3} |a_i|^2 p_i^{1-\alpha} + \mu \left(1 - \sum_{i=0}^{3} p_i\right) + \sum_{i=0}^{3} \mu_i p_i,
\end{equation}
where $\mu$ is the multiplier associated with the normalization constraint of probability, and $\mu_i \ge 0$ are the multipliers associated with the positivity constraints.\par
The KKT conditions are given by
\begin{equation}
\frac{\partial \mathcal{L}}{\partial p_i} =|a_i|^2(1-\alpha)p_i^{-\alpha} - \mu + \mu_i = 0,
\end{equation}
together with
$\mu_i \ge 0$, $p_i \ge 0$ and $ \mu_i p_i = 0$.
First, assume that $a_i \neq 0$.
Since $p_i^{-\alpha} \to \infty$ as $p_i \to 0^+$,
the optimal solution cannot be on the boundary $p_i = 0$.
Thus, we have $0 < p_i < 1,$ and by the complementary slackness condition, $\mu_i = 0.$
Then the stationarity condition reduces to
\begin{equation}
|a_i|^2(1-\alpha)p_i^{-\alpha} = \mu.
\end{equation}
which implies that the optimal weights satisfy
\begin{equation}
p_i \propto |a_i|^{2/\alpha}.
\end{equation}
Thus the closed-form solution is
\begin{equation}
p_i^{\text{opt}} = \frac{|a_i|^{2/\alpha}}{\sum_{j=0}^{3} |a_j|^{2/\alpha}}.
\end{equation}
By substituting the optimal weights $p_i^{\text{opt}}$ into the $\alpha$-affinity, the maximum value becomes
\begin{equation}
\max \mathcal{A}_\alpha = \left( \sum_{i=0}^{3} |a_i|^{2/\alpha} \right)^{\alpha}.
\end{equation}
The minimal $\alpha$-affinity Pauli distance is therefore
\begin{equation}
\widetilde{D}_{\mathcal{P}}(\mathcal{U})
= 1 - \left( \sum_{i=0}^{3} |a_i|^{2/\alpha} \right)^{\alpha}.
\end{equation}
\begin{figure*}[tb]
\centering
\begin{minipage}[t]{0.48\textwidth}
\centering
\includegraphics[width=\textwidth]{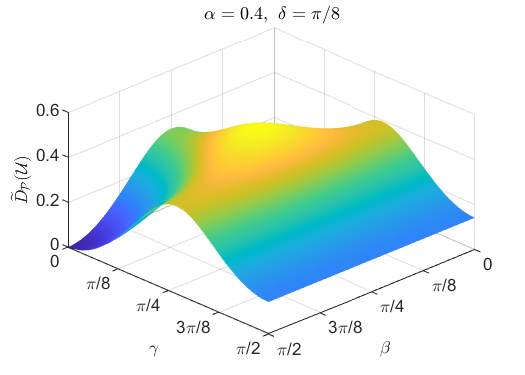}
\vspace{-0.5em}
(a)
\end{minipage}
\hfill
\begin{minipage}[t]{0.48\textwidth}
\centering
\includegraphics[width=\textwidth]{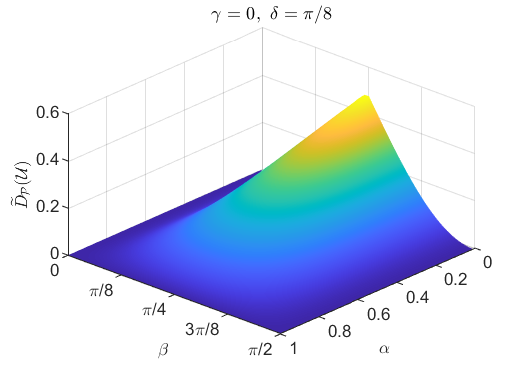}
\vspace{-0.5em}
(b)
\end{minipage}

\vspace{1em}

\begin{minipage}[t]{0.48\textwidth}
\centering
\includegraphics[width=\textwidth]{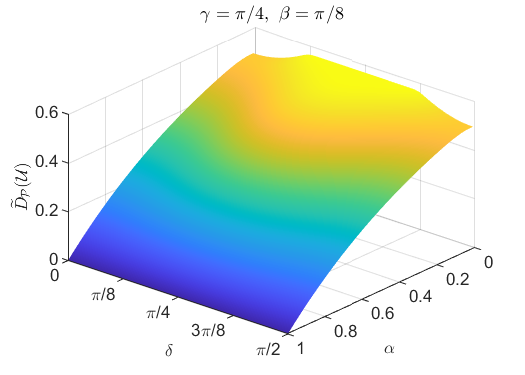}
\vspace{-0.5em}
(c)
\end{minipage}
\hfill
\begin{minipage}[t]{0.48\textwidth}
\centering
\includegraphics[width=\textwidth]{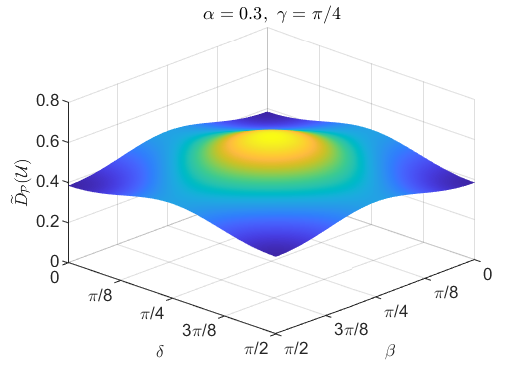}
\vspace{-0.5em}
(d)
\end{minipage}

\caption{Optimal convex approximation of a single-qubit unitary channel $\mathcal{U}(\gamma,\beta,\delta)$ with respect to the full Pauli channel set under the $\alpha$-affinity metric. The problem is solved by finding the closest Pauli channel to the target unitary channel, yielding the optimal Pauli distance $\widetilde{D}_{\mathcal{P}}[\mathcal{U}(\gamma,\beta,\delta)]$. The solution provides the optimal distance, here plotted versus different pairs of parameters under various fixed conditions: (a) varying $\beta$ and $\gamma$ for fixed $\alpha=0.4$ and $\delta=\pi/8$; (b) varying $\alpha$ and $\beta$ for fixed $\gamma=0$ and $\delta=\pi/8$; (c) varying $\alpha$ and $\delta$ for fixed $\gamma=\pi/4$ and $\beta=\pi/8$; (d) varying $\beta$ and $\delta$ for fixed $\alpha=0.3$ and $\gamma=\pi/4$.}
\label{fig:optimal_distance_pauli}
\end{figure*}
Substituting the explicit expressions
\begin{equation}
\begin{aligned}
|a_0|^2 = \cos^2\gamma\cos^2\beta, 
|a_1|^2 = \sin^2\gamma\sin^2\delta, \\
|a_2|^2 = \sin^2\gamma\cos^2\delta, 
|a_3|^2 = \cos^2\gamma\sin^2\beta,
\end{aligned}
\end{equation}
we obtain the full closed-form expression
\begin{equation}
\begin{aligned}
\widetilde{D}_{\mathcal{P}}(\mathcal{U})
= 1 - \Bigg((\cos\gamma\cos\beta)^{2/\alpha}
+ (\sin\gamma\sin\delta)^{2/\alpha} + (\sin\gamma\cos\delta)^{2/\alpha}+ (\cos\gamma\sin\beta)^{2/\alpha}\Bigg)^{\alpha}.
\end{aligned}
\end{equation}\par
If some coefficients satisfy $a_{i}=0$, the corresponding optimal probability is automatically zero:
$p_i^{\text{opt}} = 0,$ the optimization problem then reduces to the lower-dimensional subspace spanned by the nonzero coefficients and the above expressions remain valid in this case.\par
By using the full Pauli channel set for convex approximation, we obtain a more flexible and accurate approximation structure than the covariant channel family. Under the $\alpha$-affinity metric, the optimal weights are solved analytically, and the minimal distance is given in closed form. The result shows that the approximation performance depends explicitly on the unitary channel parameters and the metric parameter $\alpha$, which provides a unified and tractable criterion for evaluating the implementation of unitary operations under Pauli noise constraints.\par

Figure 2 presents the optimal $\alpha$-affinity Pauli distance $\widetilde{D}_{\mathcal{P}}(\mathcal{U})$ for the single-qubit unitary channel $\mathcal{U}(\gamma,\beta,\delta)$ under various fixed parameter conditions. Compared with the covariant channel family, the full Pauli basis provides a richer set of convex weights, allowing the approximation to adapt more closely to the structure of the target unitary channel. As a result, the minimal distance is generally reduced over a wide range of parameters, reflecting the stronger expressive capability of the complete Pauli basis.
Figure 2(a) demonstrates the optimal Pauli distance as a function of $\beta$ and $\gamma$ for fixed $\alpha=0.4$ and $\delta=\pi/8$. The surface shows a characteristic pattern of local minima and maxima, reflecting how variations in the unitary parameters modify the Bell-basis decomposition. The distance is particularly small when the channel components are well balanced, highlighting the flexibility of the Pauli channel family in approximating such symmetric configurations.
Figure 2(b) displays the minimal distance for the case $\gamma=0$ and $\delta=\pi/8$, where the unitary channel is dominated by a subset of Bell components. Here, the distance varies smoothly with $\alpha$ and $\beta$, with the smallest errors occurring for moderate values of $\alpha$ and balanced contributions from the remaining components. This shows that the Pauli basis can efficiently approximate unitary channels whose Bell-basis weights are highly concentrated, leading to small approximation errors in large parameter regions.
Figure 2(c) illustrates the optimal distance for fixed $\gamma=\pi/4$ and $\beta=\pi/8$, varying with $\alpha$ and $\delta$. The surface reveals a clear dependence on the affinity parameter $\alpha$, with the minimal distance generally decreasing as $\alpha$ approaches 1. This behavior reflects the increasing weight given to the dominant components in the $\alpha$-affinity measure, allowing the Pauli channel set to better reproduce the structure of the target unitary channel for higher $\alpha$.
Figure 2(d) shows the optimal distance for a relatively small affinity parameter $\alpha=0.3$ and fixed $\gamma=\pi/4$, varying with $\beta$ and $\delta$. At smaller $\alpha$, the $\alpha$-affinity becomes more sensitive to imbalances in the Bell-basis components, leading to a more pronounced variation in the approximation distance. The figure clearly shows that the distance grows rapidly in parameter regions where the dominant components are strongly unbalanced, demonstrating how $\alpha$ controls the geometric sensitivity of the channel distance.
\section{Pauli Distance of Amplitude Damping Channels}\label{sec5}
In this section, we investigate the optimal convex approximation of the amplitude damping channel under the Pauli channel set, based on the $\alpha$-affinity metric. Following the same optimization framework, we derive the Choi matrix of the amplitude damping channel, compute the diagonal components $q_i$ in the Bell basis, solve the optimal convex weights via the Lagrange multiplier method, and obtain the closed-form expression of the minimal $\alpha$-affinity Pauli distance.\par
The amplitude damping channel describes the energy dissipation process
of a single qubit coupled to an environment.
Its Kraus representation is given by
\begin{equation}
\Gamma(\rho)
= K_0 \rho K_0^\dagger
+ K_1 \rho K_1^\dagger,
\end{equation}
where the two Kraus operators take the form
\begin{equation}
K_0 =
\begin{pmatrix}
1 & 0 \\
0 & \sqrt{1-r}
\end{pmatrix},
\quad
K_1 =
\begin{pmatrix}
0 & \sqrt{r} \\
0 & 0
\end{pmatrix},
\end{equation}
and $r\in[0,1]$ is the damping parameter that characterizes
the decay strength from the excited state $|1\rangle$
to the ground state $|0\rangle$.  The Choi matrix of the amplitude damping channel is
\begin{equation}
R_{\Gamma}=(\Gamma\otimes I)|\eta\rangle\langle\eta|.
\end{equation}
Substituting the Kraus representation yields
\begin{equation}
R_\Gamma = \frac{1}{2}
\begin{pmatrix}
1 & 0 & 0 & \sqrt{1-r} \\
0 & r & 0 & 0 \\
0 & 0 & 0 & 0 \\
\sqrt{1-r} & 0 & 0 & 1-r
\end{pmatrix},
\end{equation}
The eigenvalues of $R_{\Gamma}$ are
$\lambda_1=\frac{2-r}{2}$, $\lambda_2=\frac{r}{2}$ and $ \lambda_3=\lambda_4=0$.
Thus, the $\alpha$-th power of $R_\Gamma$ is
\begin{equation}
\begin{aligned}
R_\Gamma^\alpha &=
\left(\frac{2-r}{2}\right)^\alpha \frac{1}{2-r}
\begin{pmatrix}
1 & 0 & 0 & \sqrt{1-r} \\
0 & 0 & 0 & 0 \\
0 & 0 & 0 & 0 \\
\sqrt{1-r} & 0 & 0 & 1-r
\end{pmatrix} 
\quad+
\left(\frac{r}{2}\right)^\alpha
\begin{pmatrix}
0 & 0 & 0 & 0 \\
0 & 1 & 0 & 0 \\
0 & 0 & 0 & 0 \\
0 & 0 & 0 & 0
\end{pmatrix}.
\end{aligned}
\end{equation}

The convex combination of Pauli channels used as the approximation set is
\begin{equation}
\mathcal{P}(\rho)
= \sum_{i=0}^{3} p_i \sigma_i \rho \sigma_i,
\end{equation}
where $p_i\ge 0$ and $\sum_{i=0}^{3} p_i = 1$,
and
$\{\sigma_0=I,
\sigma_1=\sigma_x,
\sigma_2=\sigma_y,
\sigma_3=\sigma_z  \}$
are the Pauli matrices.
The Choi matrix of the Pauli channel $\mathcal{P}$ is
\begin{equation}
R_{\mathcal{P}}=
\begin{pmatrix}
\dfrac{p_0+p_3}{2} & 0 & 0 & \dfrac{p_0-p_3}{2} \\[4pt]
0 & \dfrac{p_1+p_2}{2} & \dfrac{p_1-p_2}{2} & 0 \\[4pt]
0 & \dfrac{p_1-p_2}{2} & \dfrac{p_1+p_2}{2} & 0 \\[4pt]
\dfrac{p_0-p_3}{2} & 0 & 0 & \dfrac{p_0+p_3}{2}
\end{pmatrix}.
\end{equation}
For $0<\alpha<1$, its $(1-\alpha)$-th power is
\begin{equation}
\setlength{\arraycolsep}{4pt}
R_{\mathcal{P}}^{1-\alpha} =
\begin{pmatrix}
\frac{p_0^{1-\alpha}+p_3^{1-\alpha}}{2} & 0 & 0 & \frac{p_0^{1-\alpha}-p_3^{1-\alpha}}{2} \\
0 & \frac{p_1^{1-\alpha}+p_2^{1-\alpha}}{2} & \frac{p_1^{1-\alpha}-p_2^{1-\alpha}}{2} & 0 \\
0 & \frac{p_1^{1-\alpha}-p_2^{1-\alpha}}{2} & \frac{p_1^{1-\alpha}+p_2^{1-\alpha}}{2} & 0 \\
\frac{p_0^{1-\alpha}-p_3^{1-\alpha}}{2} & 0 & 0 & \frac{p_0^{1-\alpha}+p_3^{1-\alpha}}{2}
\end{pmatrix}.
\end{equation}
Substituting the Choi matrices into the definition of $\alpha$-affinity, we obtain
\begin{equation}
\begin{aligned}
\mathcal{A}_\alpha(\Gamma,\mathcal{P})=\mathrm{Tr}\!\left(R_{\Gamma}^{\alpha}R_{\mathcal{P}}^{1-\alpha}\right)=\sum_{i=0}^3 q_i\,p_i^{1-\alpha}.
\end{aligned}
\end{equation}
The corresponding distance is
\begin{equation}
D_\alpha(\Gamma)
= 1 - \mathcal{A}_\alpha(\Gamma,\mathcal{P})
= 1 - \sum_{i=0}^3 q_i \, p_i^{1-\alpha}.
\end{equation}
Define the diagonal coefficients
\begin{equation}
q_i = \langle B_i | R_\Gamma^\alpha | B_i \rangle,
\end{equation}
where $|B_0\rangle, |B_1\rangle, |B_2\rangle, |B_3\rangle$ are the Bell basis.
Direct calculation gives
\begin{equation}
\begin{aligned}
q_0 = \frac{\left(1-\frac{r}{2}\right)^\alpha}{2(2-r)} \left( 2 - r + 2\sqrt{1-r} \right), 
q_1 = \frac{1}{2} \left( \frac{r}{2} \right)^\alpha, \\
q_2 =\frac{1}{2} \left( \frac{r}{2} \right)^\alpha, 
q_3 = \frac{\left(1-\frac{r}{2}\right)^\alpha}{2(2-r)} \left( 2 - r - 2\sqrt{1-r} \right).
\end{aligned}
\end{equation}\par
The optimal approximation is equivalent to maximizing $\mathcal{A}_\alpha$ under the constraint $\sum_{i=0}^3 p_i=1 $ and $p_i \geq 0$. We construct the Lagrangian  function
\begin{equation}
\mathcal{L}=\sum_{i=0}^3 q_i\,p_i^{1-\alpha}+\mu\left(1-\sum_{i=0}^3 p_i\right) + \sum_{i=0}^{3} \mu_i p_i,
\end{equation}
Taking the partial derivative and setting it to zero yields
\begin{equation}
\frac{\partial\mathcal{L}}{\partial p_i}=(1-\alpha)q_i\,p_i^{-\alpha}-\mu+\mu_i=0,
\end{equation}
which implies
\begin{equation}
p_i\propto q_i^{1/\alpha}.
\end{equation}
Thus, the optimal convex weights are
\begin{equation}
p_i^{\text{opt}}=\frac{q_i^{1/\alpha}}{\sum_{j=0}^3 q_j^{1/\alpha}}.
\end{equation}
Substituting $p_i^{\text{opt}}$ into the $\alpha$-affinity, we obtain
\begin{equation}
\max\mathcal{A}_\alpha=\left(\sum_{i=0}^3 q_i^{1/\alpha}\right)^{\alpha}.
\end{equation}
The minimal $\alpha$-affinity Pauli distance of the amplitude damping channel is
\begin{equation}
\widetilde{D}_{\mathcal{P}}(\Gamma)
=1-\left(\sum_{i=0}^3 q_i^{1/\alpha}\right)^{\alpha}.
\end{equation}
Substituting the diagonal coefficients $q_0, q_1, q_2, q_3$ into the distance formula,
the explicit closed-form expression of the minimal $\alpha$-affinity Pauli distance
for the amplitude damping channel is given by
\begin{equation}
\begin{aligned}
\widetilde{D}_{\mathcal{P}}(\Gamma)
=&\ 1 - \Bigg\{ 2\left[ \frac{1}{2} \left( \frac{r}{2} \right)^\alpha \right]^{1/\alpha} + \left[ \frac{\left(1-\frac{r}{2}\right)^\alpha}{2(2-r)} \left(2 - r + 2\sqrt{1-r}\right) \right]^{1/\alpha} \\
&+ \left[ \frac{\left(1-\frac{r}{2}\right)^\alpha}{2(2-r)} \left(2 - r - 2\sqrt{1-r}\right) \right]^{1/\alpha} \Bigg\}^\alpha.
\end{aligned}
\end{equation}
\begin{figure}[tbp]
  \centering
  \includegraphics[width=0.65\textwidth]{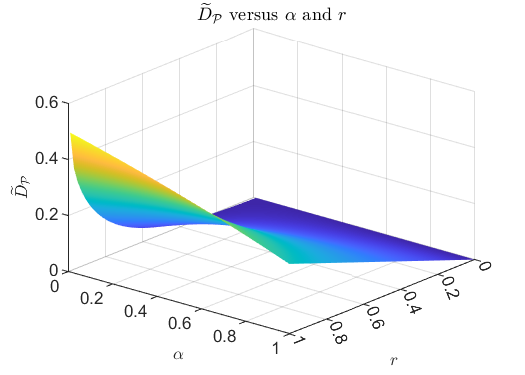}
  \caption{The optimal convex approximation distance $\widetilde{D}_{\mathcal{P}}(\Gamma)$ is presented as a function of parameters $\alpha$ and $r$, illustrating how the minimal distance changes with these variables.}
  \label{fig:untitled9}
\end{figure}
In the following, we systematically analyze the feasibility boundary of convex weights and the full-range parametric behavior for $r\in[0,1]$.

Firstly, we consider the boundary cases of convex weights $p_i\in[0,1]$.
At the endpoints $p_i=0$ and $p_i=1$, the $\alpha$-affinity becomes smaller and the corresponding channel distance is larger than that achieved at the interior critical point.
Consequently, the objective function cannot attain its minimum on the boundary of the feasible domain.
For all admissible parameters $0<\alpha<1$ and $0\le r\le 1$, the optimal convex weights $p_i^{\text{opt}}$ solved by the Lagrange multiplier method always lie strictly inside the interval $(0,1)$, demonstrating that the interior critical point is the unique global minimizer.

We further divide the parameter range $r\in[0,1]$ into three typical cases.\par
(1) When $r=0$, the amplitude damping channel degenerates into the identity channel.
The eigenvalues and diagonal elements of its Choi matrix reduce to a fixed distribution.
Substituting $r=0$ into the closed-form formula yields a constant minimal distance.
Even in this degenerate case, the optimal convex weights still remain in the interior of the feasible region without boundary collapse.

(2) The case $r=1$ corresponds to the full damping limit of the amplitude damping channel.
The spectrum of the Choi matrix degenerates obviously, and the coefficient distribution becomes highly concentrated.
Nevertheless, the optimal convex weights still stay inside $(0,1)$ rather than sticking to the boundary $p_i=0$ or $p_i=1$.
The derived closed-form distance expression is well-defined and continuous at $r=1$.

(3) For intermediate parameters $0<r<1$, the amplitude damping channel is non-degenerate with distinct eigenvalues of the Choi matrix.
The optimal convex weights and the minimal $\alpha$-affinity Pauli distance vary smoothly and continuously with $r$.
Throughout this interval, the optimal solution always locates in the interior of the simplex and never saturates to the boundary.

Overall, the closed-form solutions of optimal weights and minimal distance are valid over the entire parameter range $0\le r\le 1$.\par

Figure 3 presents the minimal $\alpha$-affinity Pauli distance $\widetilde{D}_{\mathcal{P}}(\Gamma)$ for the amplitude damping channel. The distance increases monotonically with the damping parameter $r$, indicating that stronger dissipative effects make the amplitude damping channel more difficult to approximate using Pauli channels. Moreover, different choices of the affinity parameter $\alpha$ lead to different sensitivity behaviors, showing that $\alpha$-affinity provides a tunable characterization of channel distinguishability.\par
In summary, we have analytically derived the optimal convex approximation of the amplitude damping channel over the Pauli channel set based on the $\alpha$-affinity distance. Closed-form expressions for both the optimal probability distribution and the minimal approximation distance have been obtained without numerical optimization. The results reveal explicit relations between the damping parameter, the affinity parameter $\alpha$, and the achievable approximation accuracy, thereby providing a systematic and analytically tractable approach to the study of noisy quantum channel approximation under Pauli constraints.

\section{Conclusion}\label{sec6}

In this paper, we establish a unified analytical framework for the optimal convex approximation of quantum channels based on the $\alpha$-affinity measure. We derive a channel distance metric induced by the $\alpha$-affinity and the Choi–Jamiolkowski isomorphism, and prove that it satisfies several fundamental properties required for a physically meaningful channel metric, including positivity, identity of indiscernibles, contractivity under superchannels, and joint convexity. Within this framework, the channel approximation problem is reformulated as a constrained convex optimization over normalized Choi states.\par
This reformulation enables us to derive analytical solutions for several important classes of single-qubit quantum channels. Specifically, we obtain the optimal convex approximation of unitary channels within the SU(2)-covariant channel family and the complete Pauli channel set, as well as the optimal Pauli approximation of amplitude-damping channels. In all cases, the optimal convex coefficients and the corresponding minimal $\alpha$-affinity distances are obtained in closed form.\par
Our results demonstrate that the $\alpha$-affinity is not only a mathematically tractable distance measure, but also provides an operationally meaningful framework for studying the practical implementation of quantum channels under limited experimental resources. Compared with conventional channel distances such as the diamond norm, the proposed approach admits explicit analytical optimization and reveals direct relations between channel parameters, noise structures, and approximation performance. The framework developed here can be naturally extended to higher-dimensional systems, general noisy channels, and other classes of free operations in quantum resource theories. It may also find useful applications in quantum error mitigation, noisy gate synthesis, channel simulation, and fault-tolerant quantum computation.

\bmhead{Acknowledgements}

Project supported by the Fundamental Research Projects of Shanxi Province (Grant No. 202203021222225 and No.
202203021211260), the National Natural Science Foundation of China (Grant Nos. 12175029, 12011530014, and 11775040), and the Key Research and Development Project of Liaoning Province (Grant No. 2020JH2/10500003).
\section*{Declarations}

\subsection*{Funding}
This work was supported by the Fundamental Research Projects of Shanxi Province (Grant No. 202203021222225 and No.
202203021211260), the National Natural Science Foundation of China (Grant Nos. 12175029, 12011530014, and 11775040), and the Key Research and Development Project of Liaoning Province (Grant No. 2020JH2/10500003).

\subsection*{Conflict of interest}
The authors declare that they have no conflict of interest relevant to this study.

\subsection*{Ethical approval and consent to participate}
Not applicable. This is a theoretical study and does not involve human or animal subjects.

\subsection*{Consent for publication}
All authors have reviewed and approved the manuscript for publication.

\subsection*{Data availability}
Not applicable. All theoretical derivations and results are included in the main text and appendices.

\subsection*{Materials availability}
Not applicable.

\subsection*{Code availability}
Not applicable. No custom code or algorithm was used to generate the results of this study.

\subsection*{Author contribution}
Liqiang Zhang contributed to conceptualization, methodology, model development, experiments, critical revision of the manuscript, data analysis and project administration. Chengling Fu contributed to manuscript drafting, and supervision. Liuyong Cheng contributed to technical support, code review, and contribution to result interpretation. Guohui Yang contributed to technical support, code review, and contribution to result interpretation. Changshui Yu contributed to technical support, code review, and contribution to result interpretation.


\end{document}